\documentclass{raa}

\usepackage{graphicx,times}
\usepackage{natbib}
\usepackage{amssymb,amsmath}
\bibpunct{(}{)}{;}{a}{}{,}
\usepackage{placeins} %%自行添加为了使图片不跑到别的章节

\usepackage[pagebackref=false]{hyperref}

\begin{document}

   \title{Timing and Spectral Studies of PSR J2022+3842 with NICER and NuSTAR}

 \volnopage{ {\bf 20XX} Vol.\ {\bf X} No. {\bf XX}, 000--000}
   \setcounter{page}{1}

   \author{Jia-Ning Hu %% Put your Chinese name in "( )" if you like. Note to open line 11 "\usepackage[UTF8]{ctex}"
      \inst{1}
   \and Xiang-Hua Li\thanks{Corresponding author: xhli@ynu.edu.cn}
      \inst{1}
   \and Xian-Ao Wang
      \inst{1}
   \and Han-Long Peng
      \inst{3}
   \and Shi-Qi Zhou
      \inst{4}
   \and Wen-Tao Ye
      \inst{2}
   \and Shi-Jie Zheng
      \inst{2}
   \and Ze-Jun Jiang
      \inst{1}
   \and Ming-Yu Ge\thanks{Corresponding author: gemy@mail.ihep.ac.cn}
      \inst{2,5} 
   }

   \institute{Department of Astronomy, Yunnan University, Kunming 650090, People’s Republic of China; {\it hujianing@stu.ynu.edu.cn}\\
        \and
             State Key Laboratory of Particle Astrophysics, Institute of High Energy Physics, Chinese Academy of Sciences, Beijing 100049, People’s Republic of China\\
        \and
             Department of Physics and Institute of Theoretical Physics, Nanjing Normal University, Nanjing, 210023, Jiangsu, People’s Republic of China\\
        \and
             School of Physics and Astronomy, China West Normal University, Nanchong 637002, People’s Republic of China\\
        \and
             University of Chinese Academy of Sciences, Chinese Academy of Sciences, Beijing 100049, People’s Republic of China\\
%\vs \no
%   {\small Received 20XX Month Day; accepted 20XX Month Day}
}

\abstract{We report on the long-term timing analysis of PSR J2022+3842 using observations from the Neutron Star Interior Composition Explorer (NICER), along with spectral properties derived from joint observations with NICER and the Nuclear Spectroscopic Telescope Array (NuSTAR). Two large glitches are identified around MJD 58335 with $\mathrm{\Delta\nu}=25.35(2)\times10^{-6}~\text{Hz}$ and MJD 58875 with $\mathrm{\Delta\nu}=52.078(6)\times10^{-6}~\text{Hz}$ . Furthermore, phase-resolved spectroscopy reveals that the X-ray emission is well described by a power-law model across different phase intervals. The phase-integrated X-ray spectrum (1--79 keV) has a photon index of $\mathrm{\Gamma}=1.22(7)$, yielding an unabsorbed 0.5--10 keV flux of $8.9(6)\times10^{-13}\ \text{erg}~\text{cm}^{-2}~\text{s}^{-1}$. The main pulse spectrum (1.2--79 keV) and the inter-pulse spectrum (1--70 keV) are harder with $\mathrm{\Gamma}=1.17(4)$ and $\mathrm{\Gamma}=1.03^{+0.07}_{-0.06}$ separately, producing an unabsorbed 0.5--10 keV flux of $33.2(2)\times10^{-13}\ \text{erg}~\text{cm}^{-2}~\text{s}^{-1}$ and $29(3)\times10^{-13}\ \text{erg}~\text{cm}^{-2}~\text{s}^{-1}$. Investigation of the pulse profile evolution with time shows that no significant variations were observed.
}

   \authorrunning{J.-N. Hu et al. }            %author_head in even pages
   \titlerunning{Timing and Spectral Studies of PSR J2022+3842}  % title_head in odd pages
   \maketitle

%________________________________________________ sections below
% 
\section{Introduction}           %% first-level sections will be auto-capitalized
\label{sect:intro}

Pulsars, as highly magnetized and rapidly rotating neutron stars, serve as unique laboratories for studying matter states, photon acceleration, and gravitational waves under extreme physical conditions. Their long-term frequency evolution is so stable that it can be compared to an atomic clock (\citealt{1987Sci...238..761R, 2015HiA....16..207H}). However, perfect periodic signals often exhibit irregularities, primarily due to two main causes: apparent timing noise over long timescales and sudden change in spin frequency, referred to as a glitch. The timing noise in young pulsars with high spin-down luminosity is particularly significant and complex (\citealt{2019MNRAS.489.3810P}). Additionally, the observed state transition behaviors in young pulsars can also influence pulsar timing (\citealt{2020ApJ...900L...7G}).
These two phenomena challenge the precise determination of fundamental pulsar parameters and the detection of subtle physical effects such as small-amplitude glitches.

Glitches are characteristic of young pulsars and provide valuable insight into the dynamics of superfluid within neutron star interiors, as well as the crust-core coupling mechanism (\citealt{2014ApJ...788L..11G}). Based on the magnitude of the frequency change relative to the frequency prior to the glitch, the glitches can be classified into two groups: large size glitches with $\mathrm{\delta\nu/\nu}>10^{-7}$ and small size glitches with $\mathrm{\delta\nu/\nu}<10^{-7}$ (\citealt{2019RAA....19...89O}). According to the temporal characteristics of frequency and 
the frequency derivative variations, glitches can be classified into four categories (\citealt{2022Univ....8..641Z, 2015IJMPD..2430008H}): normal glitches, slow glitches, glitches with delayed spin-ups (\citealt{2020ApJ...896...55G}), and anti-glitches (\citealt{2024ApJ...967L..13T}). Typically, large glitches are explained by the vortex creep model, whereas small glitches are accounted for by the crustquake model (\citealt{2022Univ....8..641Z}).

Timing noise is also a characteristic feature of young pulsars. \citealt{1981ApJ...245.1060C} first proposed possible theoretical models of timing noise in isolated pulsars. Then \citealt{1989ASIC..262..503C} proposed that timing noise is the result of magnetospheric fluctuations and internal superfluid unpinning. Investigating the correlation between timing noise, frequency derivative and pulse profile, \citealt{2010Sci...329..408L} found observational evidence that timing noise comes from changes in magnetospheric state. Besides, \citealt{2010ApJ...725.1607S} presented the function for the timing noise power spectrum of different kinds of pulsar, allowing the development of models to mitigate the impact of timing noise on the detection of gravitational waves. \citealt{2021MNRAS.502..478G} optimized the white and red noise model and built the basis for numerical calculation of the timing residuals caused by timing noise.

PSR J2022+3842 is a young Crab-like pulsar with high spin-down luminosity and low X-ray conversion efficiency. Young pulsars like PSR J2022+3842 exhibit active glitch behavior, and long-term timing observations of such objects can help us understand pulsar interior structures, as well as the storage and transfer mechanisms of angular momentum.

PSR J2022+3842 was first discovered in supernova remnant G76.9+1.0 in 2010 (\citealt{2010cosp...38.2807A}). It has been observed in radio (\citealt{2010cosp...38.2807A}), X-ray (\citealt{2010cosp...38.2807A, 2011ApJ...739...39A, 2014ApJ...790..103A}) and gamma-ray bands (\citealt{2011ATel.3466....1P, 2022PhDT........16L}). In \citealt{2014ApJ...790..103A}, the spin-down power of PSR J2022+3842 was calculated as $\mathrm{\dot{E}} = 3.0\times10^{37}~\text{erg}~\text{s}^{-1}$ with the magnetic field $\mathrm{B} = 2.1\times10^{12}~\text{G}$. The pulsar's spectrum can be well fitted with a hard power-law (PL) model with the photon index $\mathrm{\Gamma} = 0.9 \pm 0.1$. Only one large glitch was reported around MJD 54680 (\citealt{2011ApJ...739...39A}). 
We used the timing analysis result in \citealt{2022PhDT........16L} as initial parameters to perform a detailed timing analysis, and no proper motion was detected. Unlike Vela and Vela-like pulsars, no pulse profile variability occurred during both glitches, indicating different glitch mechanisms in different pulsar populations (\citealt{2016ApJ...820...64P, 2014ApJ...790..103A, 2024MNRAS.533.4274L}). Besides, we found a glitch-like event arises from the discontinuity in ephemeris caused 
by timing noise around MJD 59260. We used the method proposed in \citealt{2024PASA...41..102G} to find it.
Additionally, we performed a joint spectral fit using NICER and NuSTAR data, finding that the pulsar's spectra are softer compared with \citealt{2014ApJ...790..103A}.

A comprehensive analysis of PSR J2022+3842 has been performed using NICER and NuSTAR observations. We presented the long-term evolution of PSR J2022+3842 over the past eight years. This paper is organized as follows. In Section~\ref{sec:obs}, we describe the observations and data reduction procedures for both telescopes. The results of the timing and spectral analysis derived from the procedures in Section~\ref{sec:obs} are presented in Section~\ref{sec:results}. Finally, in Section~\ref{sec:discuss}, we summarize all our findings and discuss the glitch-like event attributed to timing noise.

% Authors can give a citation as `\citealt{Michel+etal+1992}'.
% You may also use \cite, \citep and \citet for citation, and use Table~1
% or Figure~1 and so forth. Using \ref and \label for cross-references of
% Tables/Figures is a good way in adjusting/adding/removing text, tables or
% figures.

\section{Data Reduction} \label{sec:obs}

\subsection{NICER Observations} \label{subsec:nicer}

The Neutron Star Interior Composition Explorer (NICER) is a highly sensitive X-ray timing instrument with sub-microsecond timing resolution, covering the 0.2--12~keV energy range. The pulsar timing data of PSR~J2022+3842 utilized NICER observations between 2017 November and 2025 June (MJD~58000--60500), accumulating a total exposure of approximately 609~ks. We used all NICER data between this period and processed them with HEASoft (v6.34) from the NASA High Energy Astrophysics Science Archive Research Center (HEASARC 2014) and the NICER Data Analysis Software with the CALDB release  xti20240206. Because of NICER's imaging limitation, we analyzed all data from an extraction region including the pulsar and its Pulsar Wind Nebula (PWN). For data analysis:

(1) All NICER observations were preliminarily processed using the {\tt nicerl2} command with default parameters. 

(2) Observations with no Good Time Intervals (GTIs) were discarded, a situation that occurs when the exposure time is too short to produce any GTIs. 

(3) The {\tt nicerl3-lc} command was used to extract 10–15 keV light curves with a time bin size of 50 s to identify abnormally high count rates at the beginning of GTIs. Since no flare events were reported, these anomalies were attributed to instrumental effects.

(4) To mitigate the effect of these abnormally high count rates, the {\tt maketime} command was used to filter GTIs with a count rate below $\mathrm{0.5~counts~\text{s}^{-1}}$ in the 10--15~keV light curve and then {\tt nicerl2} was rerun with {\tt gtifiles}. 

\subsection{NuSTAR Observations} \label{subsec:nustar}
The Nuclear Spectroscopic Telescope Array (NuSTAR) comprises two co-aligned modules (FPMA and FPMB) and is the first hard X-ray focusing telescope sensitive in the 3--79~keV band. PSR~J2022+3842 was observed once with NuSTAR on 2018 January 15 (ObsID: 30364002002). Due to NuSTAR's angular resolution of 18\arcsec\ Full Width at Half Maximum (FWHM), we were unable to extract the pulsar and PWN regions separately. Therefore, we extracted a circular region of 70\arcsec\ radius centered on the pulsar's position (\text{ra = 20:22:21.689} \text{dec = 38:42:14.82}, \citealt{2022PhDT........16L}) as the source region and a source-free circular region of 100\arcsec\ radius as the background region. The same extraction radius was applied to both FPMA and FPMB to generate event files, and we used the {\tt nupipeline} and {\tt nuproducts} tasks to produce spectra. The NuSTAR Data Analysis Software with CALDB  v20240916 was used for analysis.

\subsection{Timing Analysis Method} \label{subsec:timing}
First of all, we ran the {\tt barycorr} task for both NICER and NuSTAR data to correct the photon arrival times to the Barycentric Dynamical Time (TDB) standard using the JPL DE405 solar system ephemeris. We used a Taylor expansion up to the second frequency derivative to compute the phase of barycenter-corrected event times:

\begin{equation}
\mathrm{\phi(t) = \phi_0 + \nu_0(t-t_0) + \frac{1}{2}\dot{\nu}_0(t-t_0)^2
          + \frac{1}{6}\ddot{\nu}_0(t-t_0)^3.}
\end{equation}
where $\mathrm{\phi_0}$ is the phase at the reference time $\mathrm{t_0}$, and $\mathrm{\nu_0}$, $\mathrm{\dot{\nu}_0}$, and $\mathrm{\ddot{\nu}_0}$ are the spin frequency and its first and second derivatives.

In timing analysis, we used only the NICER data due to a 0.09 phase difference (around 4.37~ms) between the pulse times of arrival (ToAs) of the NICER and NuSTAR data during the same time span. Including NuSTAR data would worsen the NICER timing results. \citealt{2021ApJ...908..184B} regard this difference as unmodeled measurement noise or transient processing delays during ground station clock offset measurements, not a result of intrinsic properties of PSR J2022+3842.

Due to the large timing noise, we grouped at least one observation in the analysis. For those with poor statistics, we merged them into one group until we got a pulse profile with signal-to-noise ratio higher than 8.  
Using the epoch-folding method and Pearson's $\mathrm{\chi}^{2}$ test, we searched for the frequency and its first derivative between MJD~58000 and MJD~60500.  
The epoch-folding results were fitted with a linear function $\mathrm{\nu = \nu_{0} + \dot{\nu_{0}}(t - t_{0})}$ to produce an initial timing solution. Finally, we obtained the ToAs following the same procedure described in \citeauthor{2025ApJ...992....5W}.

Around MJD 59260, the timing noise leads to a discontinuity in the ephemeris, which appears as a glitch like feature. We separately constructed the timing noise model for the timing residuals before and after the glitch-like event. Firstly, we used the pulsar module of the ENTERPRISE package (Enhanced Numerical Toolbox Enabling a Robust PulsaR Inference SuitE, imported as enterprise in Python) to load ephemerides and ToAs for the segments between MJD 58875--59260 and MJD 59260--59900 in ENTERPRISE (\citealt{2020zndo...4059815E}). Then the DYNESTY Python library was used to determine the best noise model for both segments. The Bayes factor (BF) was calculated with Jeffrey’s scale and Occam’s razor applied to select the best noise model. After identifying the best model for both segments, Parallel Tempering Markov Chain Monte Carlo (PTMCMC) was used to obtain the best-fit parameter values. Finally, we extracted residuals caused by timing noise from all timing residuals and find out if there is a true glitch event in the ‘clean’ timing residuals.

\subsection{The Phase-resolved Spectra} \label{subsec:phase_spec}
The source is faint, so using the 3C50 background directly generated by {\tt nicerl3-spec} and {\tt nuproducts} would significantly underestimate the background count rate below 2~keV. Therefore, we performed phase-resolved spectroscopy and used the off-pulse spectrum as the background for all spectra. Based on the NICER count rate histogram with 100 phase bins, we 
identified the main‑pulse and inter‑pulse phase bins as those bins whose count rate exceeds the maximum count rate in the off‑pulse region. The main pulse phase range is 0.51--0.63, the inter-pulse phase range is 0.01--0.09, and the off-pulse phase range is 0.09--0.51 and 0.63--1.00. For each segment, we defined the GTIs and extracted spectra from the newly selected data.

We used {\tt niobsmerge} to generate the NICER phase-integrated spectrum and regrouped it using {\tt grppha}. 
For phase-integrated spectra, the 1--10~keV NICER data were regrouped with a minimum of 50000 counts per bin, and the 3--79 ~keV NuSTAR data were regrouped with a minimum of 100 counts per bin.
For the main pulse spectra, the 1.2--8~keV NICER data were regrouped with a minimum of 3000 counts per bin, and the 3--79~keV NuSTAR data were regrouped with a minimum of 15 counts per bin.
For inter-pulse spectra, the 1--10~keV NICER data were regrouped with a minimum of 5000 counts per bin, and the 3--70~keV NuSTAR data were regrouped with a minimum of 15 counts per bin.
All energy bands were selected to ensure that the source counts exceeded the background.

Due to the extremely high background count rate in the NICER spectrum, we fixed the NuSTAR FPMA spectral constant parameter to 1 and compared all other spectral normalization constants to it.
Firstly, we fitted the phase-integrated spectrum with the constant*tbabs*powerlaw model. Then we fitted the main pulse and the inter-pulse spectra with the hydrogen column density $\mathrm{N_{\rm H}}$ fixed to the phase-integrated spectrum's best-fit value.

\section{Results} \label{sec:results}

\subsection{Timing Results} \label{subsec:timing_results}

\subsubsection{The Pulse Profiles} \label{subsec:profile}

\begin{figure}[htbp]
    \centering
    \includegraphics[width=1.0\textwidth]{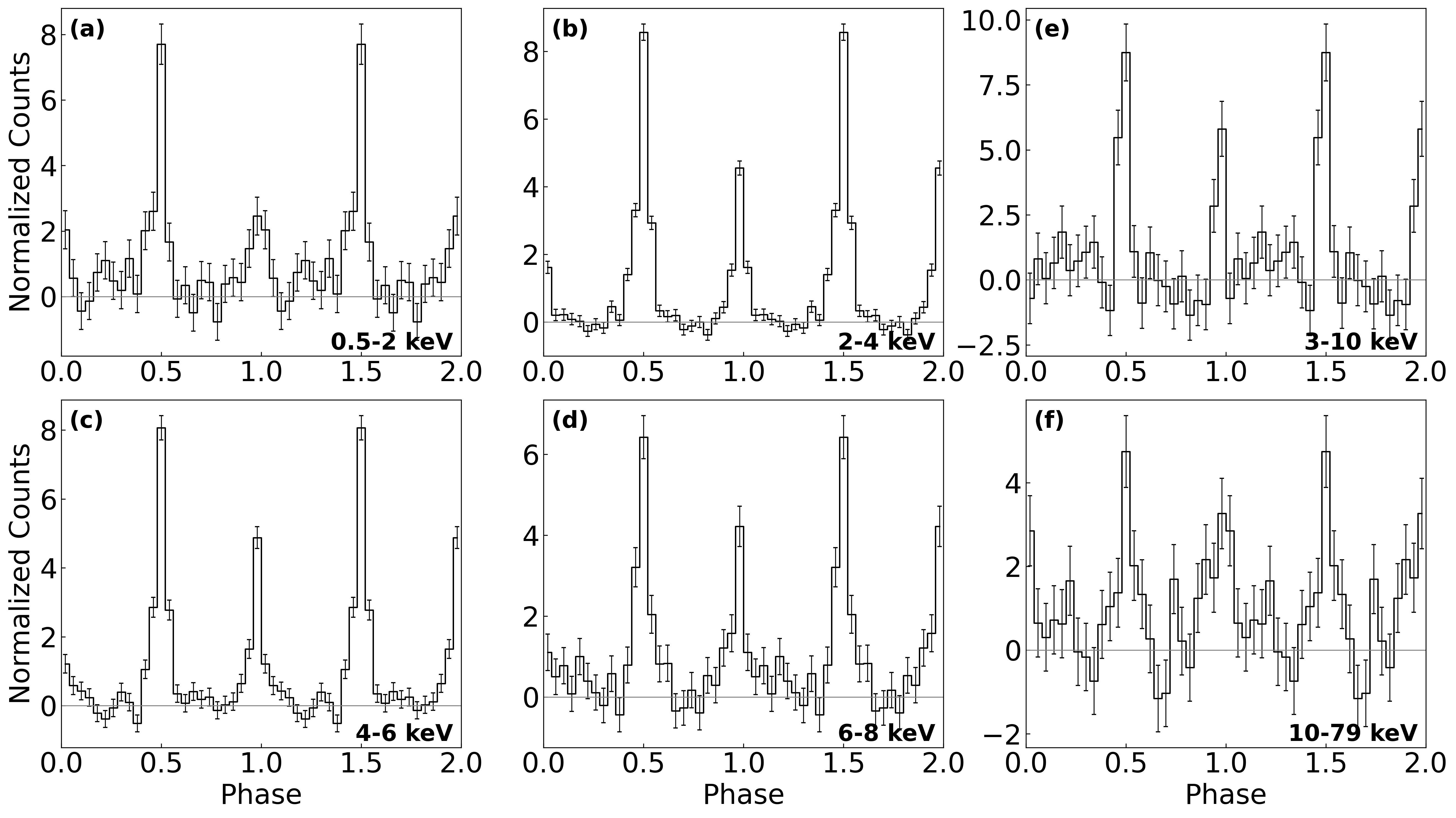}
    \caption{Pulse profiles observed by NICER and NuSTAR. Panel (a) shows the NICER 0.5--2~keV profile, Panel (b) shows the NICER 2--4~keV profile, Panel (c) shows the NICER 4--6~keV profile, Panel (d) shows the NICER 6--8~keV profile and Panel (e) shows the NuSTAR 3--10~keV profile, Panel (f) shows the NuSTAR 10--79~keV profile. All profiles include two cycles and are divided into 25 phase bins. The normalized counts are obtained by subtracting the average of the off-pulse portion from all data and then dividing by the average of the remaining portion.}
    \label{fig:profile1}
\end{figure}

\begin{figure}%[htbp]
    %\centering
    %\includegraphics[width=12.0cm, angle=0]
    \centering
    \includegraphics[width=1.0\textwidth]{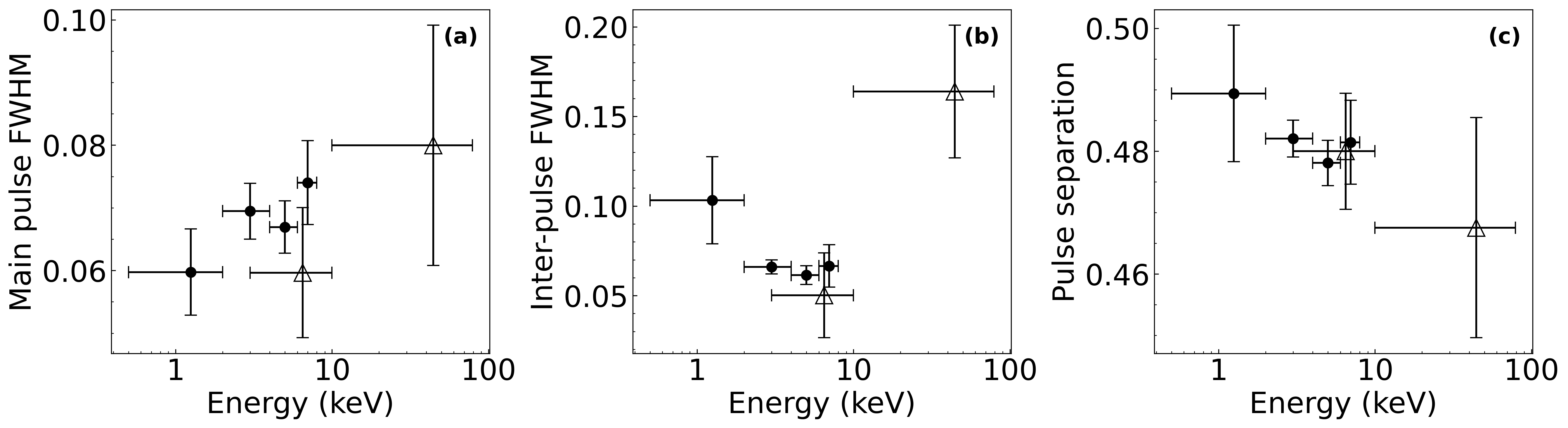}
    \caption{Energy dependence of pulse profile parameters with NICER (filled black circle) and NuSTAR (open black triangle) data. Panel (a) shows the energy dependence of main pulse FWHM, Panel (b) shows the energy dependence with the inter-pulse FWHM and Panel (c) shows the energy dependence of the phase separation between main pulse and inter-pulse. The x-axis is in log scale and each data presents an energy band in Figure~\ref{fig:profile1}.}
    \label{fig:profile2}
\end{figure}

Based on the timing results obtained, we divided all NICER data into five segments: MJD 58000--58300, MJD 58300--58875, MJD 58875--59260, MJD 59260--59900, and MJD 59900--60500. Detailed timing results are given in the following part of this paper. For pulse profile analysis, we selected four energy bands (0.5--2~keV, 2--4~keV, 4--6~keV and 6--8~keV) then stacked and aligned data from all NICER observations to obtain the total pulse profile. NuSTAR data were folded using the NICER timing solution of MJD 58000--58300. For the NuSTAR pulse profile, we selected two energy bands (3--10~keV and 10--79~keV) and processed the data the same way. The NICER and NuSTAR profiles are shown in Figure \ref{fig:profile1}. The energy dependence of the pulse profile parameters is presented in Figure \ref{fig:profile2}.

The NuSTAR pulse profiles exhibit a lower signal-to-noise ratio compared to those of NICER, likely due to an uncalibrated offset in their time-of-arrival measurements (see Section~\ref{subsec:timing}) and a lower signal-to-noise ratio due to limited exposure time. No time-dependent variation in the pulse profile was found during this work, which is more akin to the Crab Pulsar than to the Vela Pulsar, suggesting that no magnetic field rearrangement or reorganization of the overall magnetic layer structure occurred during both glitches (\citealt{2022ApJ...932...11Z, 2019NatAs...3.1122G}). For energy dependence, the main pulse FWHM is consistent across energies within $1\sigma$ errors, whereas the inter-pulse FWHM for NICER (0.5--2~keV) and NuSTAR (10--79~keV) deviates from other bands. Only the NICER 0.5--2~keV inter-pulse profile is lower and broader (Fig.~\ref{fig:profile1}), likely due to a photon index difference between main and inter-pulse spectra rather than the absorption effect, whereas the NuSTAR anomaly may stem from its low signal-to-noise ratio. 
Although the pulse separation is consistent within the $1\sigma$ uncertainty, best-fit values show a decreasing trend with increasing energy. This phenomenon has also been observed in the Crab Pulsar and can be explained by the two-gap outer gap model (\citealt{1997ApJ...476..281E}).

\subsubsection{Two Large Glitches} \label{subsec:glitch}
\begin{figure}[htbp]
    \centering
    \includegraphics[width=1.0\textwidth]{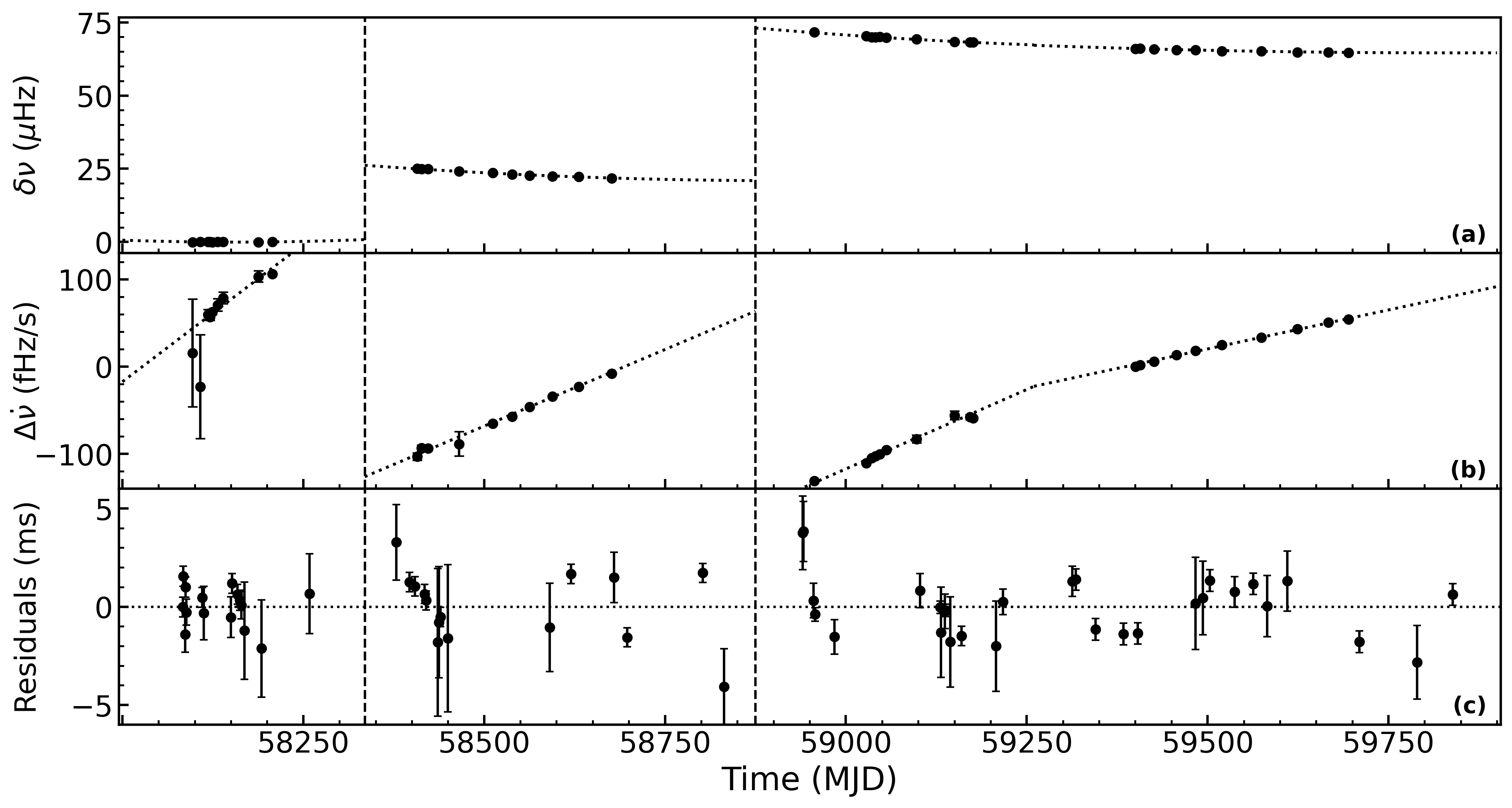}
    \caption{Step-timing analysis of PSR J2022+3842 using NICER data. The vertical black dashed lines indicate the glitch epochs at MJD 58335 and MJD 58875. The black dotted lines in Panel (a) and (b) represent the theoretical value of frequency and frequency derivative variations over time given by each segment ephemeris, with the pre-glitch timing model subtracted. Panel (a) shows the frequency residuals after subtracting the pre-glitch timing model. Panel (b) shows the evolution of $\mathrm{\dot{\nu}}$. $\mathrm{\Delta{\dot{\nu}}}$ is defined as $\mathrm{\dot{\nu}\, +\, 36514\times10^{-15}}$. Panel (c) shows the best-fit timing residuals during MJD 58000--59900. The timing residuals were obtained by concatenating the residuals resulting from fitting the TOAs in each time segment with its respective ephemeris.}
    \label{fig:glitch1}
\end{figure}

We used the method described in Section \ref{subsec:timing} and primary pulsar parameters from \citealt{2022PhDT........16L} to derive all five ephemerides. The parameters are shown in Table \ref{tab:ephemeris}.
Two large glitches occurred during MJD 58000 to MJD 59900. The glitch epochs were assigned to the midpoint between the last pre-glitch ToA and the first post-glitch ToA. Due to a data gap of approximately 100 days around both glitches, no exponential decay component was observed after either glitch. We used the tempo2 software package to obtain the final glitch parameters, which are presented in Table \ref{tab:glitch}.
The stepwise timing analysis between MJD 58000 and MJD 59900 is shown in Figure~\ref{fig:glitch1}.

Despite the two identified large glitches, we can infer that a large glitch occurred between MJD 59900 and MJD 60200 based on the frequency differences extrapolated from segments 4 and 5. However, because of the limited observations, we can only obtain three ToAs during segment 5, so it is impossible to get the parameters of this glitch.
Additionally, due to the significant timing noise caused by the large glitch in MJD 58875, data before and after MJD 59260 cannot be described by a single ephemeris. This can be visually confirmed by inspecting the shape of the timing residuals. We also used the method described in \citealt{2024PASA...41..102G} to estimate the timing noise and obtain the whitened residuals. These results are presented in Section \ref{sec:discuss}.

PSR J2022+3842 is a young pulsar with a characteristic age of 8.9 kyr (\citealt{2011ApJ...739...39A}), and the high timing noise makes it difficult to detect small glitches. Up to now, all three identified glitches and the infered glitch are all large glitches. The two large glitches identified in our work are both giltches with $\mathrm{\Delta\nu>0}$ and $\mathrm{\Delta\dot{\nu}<0}$. Using the detection boundaries defined in \citealt{2024A&A...689A.191Z}, the minimum $\mathrm{\Delta\nu}$ of normal glitches that can be detected in PSR J2022+3842 is $\mathrm{6.75\times10^{-8}Hz}$. 
We also calculated the activity parameter $\mathrm{A_{g}\approx7\times 10^{-7}yr^{-1}}$, taking small glitches affected by the timing noise into account, this value should be larger. The Crab-like 8.9~kyr pulsar PSR J2022+3842 has an activity parameter similar to the Vela Pulsar ($\mathrm{A_{g}\approx8\times 10^{-7}yr^{-1}}$, the Crab Pulsar $\mathrm{A_{g}\approx0.1\times 10^{-7}yr^{-1}}$, \citealt{2005ApJ...633.1095L}), indicating that it is cooling and may be in the process of transferring to a Vela-like pulsar from the Crab-like pulsar. 

\begin{table}
\bc
\begin{minipage}[]{1.0\textwidth}
\caption[]{Rotation parameters of PSR J2022+3842 obtained from different segments\label{tab:ephemeris}}\end{minipage}
\setlength{\tabcolsep}{5pt}
\small
 \begin{tabular}{cccccc}
  \hline%\noalign{\smallskip}
Segment& Data span& Epoch (MJD)& $\nu$ (Hz)& $\dot{\nu}$ ($\times 10^{-11}$ Hz s$^{-1}$)& $\ddot{\nu}$ ($\times 10^{-21}$ Hz s$^{-2}$)\\
  \hline\noalign{\smallskip}
1 & 58000--58300 & 58168.79 & $20.577323183(6)$  & $-3.64252(9)$ & $7.3(6)$ \\
2 & 58300--58875 & 58417.04 & $20.576566593(5)$  & $-3.66113(9)$ & $4.06(6)$ \\
3 & 58875--59260 & 58955.61 & $20.57491789(1)$  & $-3.6647(2)$ & $4.2(2)$ \\
4 & 59260--59900 & 59220.68 & $20.574079555(9)$  & $-3.65436(7)$ & $2.07(2)$ \\
5 & 59900--60500 & 60218.14 & $20.570961997(3)$  & $-3.5024(-)$ & $(-)$ \\
  \noalign{\smallskip}\hline
\end{tabular}
\ec
%% place \tablecomments and \tablerefs below \end{center| and \end{center}:
%% you may leave the table-width parameter to editors or set to your actual size
\tablecomments{0.86\textwidth}{Table 1 shows the ephemerides parameters. We only fitted the $\mathrm{\nu}$ for segment 5 and fixed $\mathrm{\dot{\nu}}$ to the value obtained from the EPOCH folding method due to the lack of data.}
\end{table}

%%%%%%%%%%%%%%%%%%%%%%%%%%%%%%%%%%%%%%%%%%%%%%%%
\begin{table}
\bc
\begin{minipage}[]{100mm}
\caption[]{Glitch parameters of PSR J2022+3842 \label{tab:glitch}}\end{minipage}
\setlength{\tabcolsep}{5pt}
\small
 \begin{tabular}{ccccc}
  \hline%\noalign{\smallskip}
Glitch epoch (MJD)& $\Delta\Phi$& $\Delta\nu_g$ ($\times 10^{-5}$ Hz)& $\Delta\dot{\nu}_g$ ($\times 10^{-13}$ Hz s$^{-1}$)& $\Delta\ddot{\nu}_g$ ($\times 10^{-21}$ Hz s$^{-2}$)\\
  \hline\noalign{\smallskip}
58335 & $-0.1(2)$ & $2.535(2)$  & $-3.19(2)$ & $-3.22(4)$ \\
58875 & $+0.6(1)$ & $5.2078(6)$  & $-2.23(2)$ & $-$ \\
  \noalign{\smallskip}\hline
\end{tabular}
\ec
%% place \tablecomments and \tablerefs below \end{center| and \end{center}:
%% you may leave the table-width parameter to editors or set to your actual size
%\tablecomments{0.86\textwidth}{Table 2 shows the glitch parameters.}
\end{table}
\FloatBarrier

\subsection{Joint Spectral Fitting Results} \label{subsec:spec_results}
\begin{figure}[htbp]
    \centering
    \includegraphics[width=0.8\textwidth]{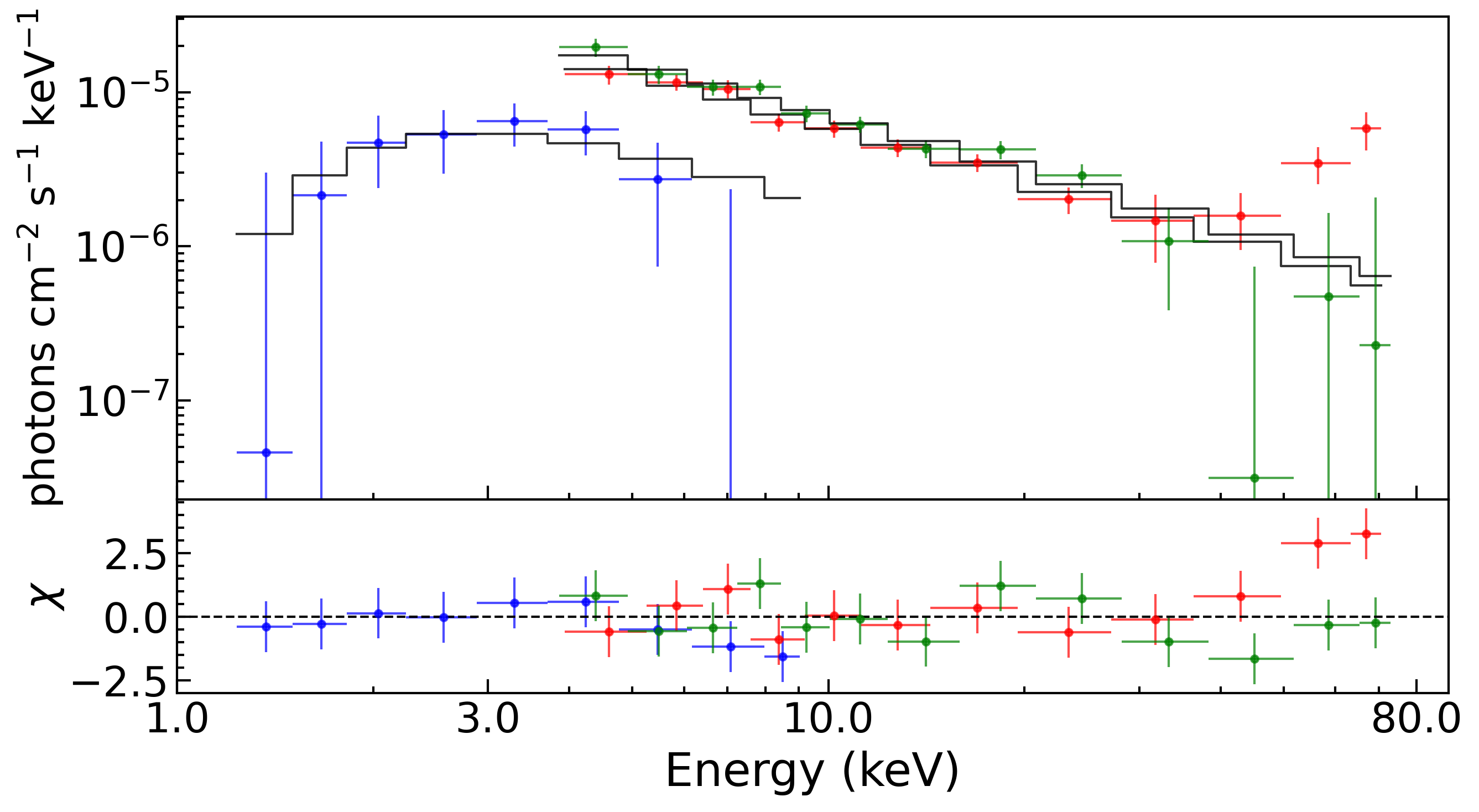}
    \caption{Phase-integrated spectra observed by NuSTAR and NICER. The top panel shows the joint spectral fit of NuSTAR and NICER data, while the bottom panel presents the residuals. The blue crosses represent the NICER spectrum in the 1--10~keV energy range, the red and green crosses correspond to the NuSTAR FPMA and FPMB spectra in the 3--79~keV range respectively. The black solid lines represent the best-fit model curves for each instrument. All energy bands are selected to ensure that the source count rate is higher than the background.}
    \label{fig:spectra1}
\end{figure}

\begin{figure}[htbp]
    \centering
    % 子图 (a)
    \begin{subfigure}[t]{0.48\textwidth}
        \centering
        \includegraphics[width=7.5cm, height=6.5cm, keepaspectratio]{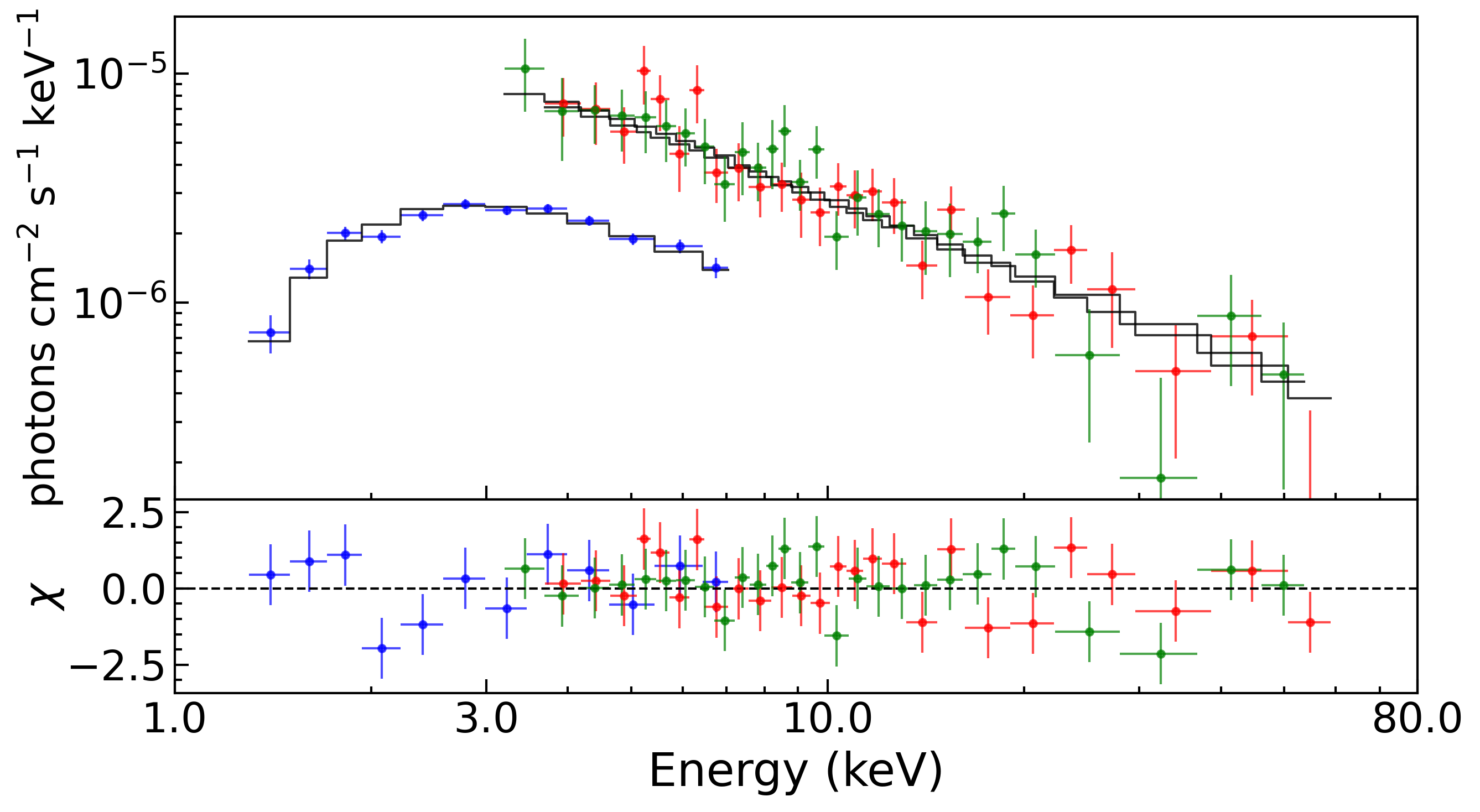}
        \caption{Main pulse spectra observed by NuSTAR and NICER. }
        \label{fig:spectra2_a}
    \end{subfigure}
    \hfill
    % 子图 (b)
    \begin{subfigure}[t]{0.48\textwidth}
        \centering
        \includegraphics[width=7.5cm, height=6.5cm, keepaspectratio]{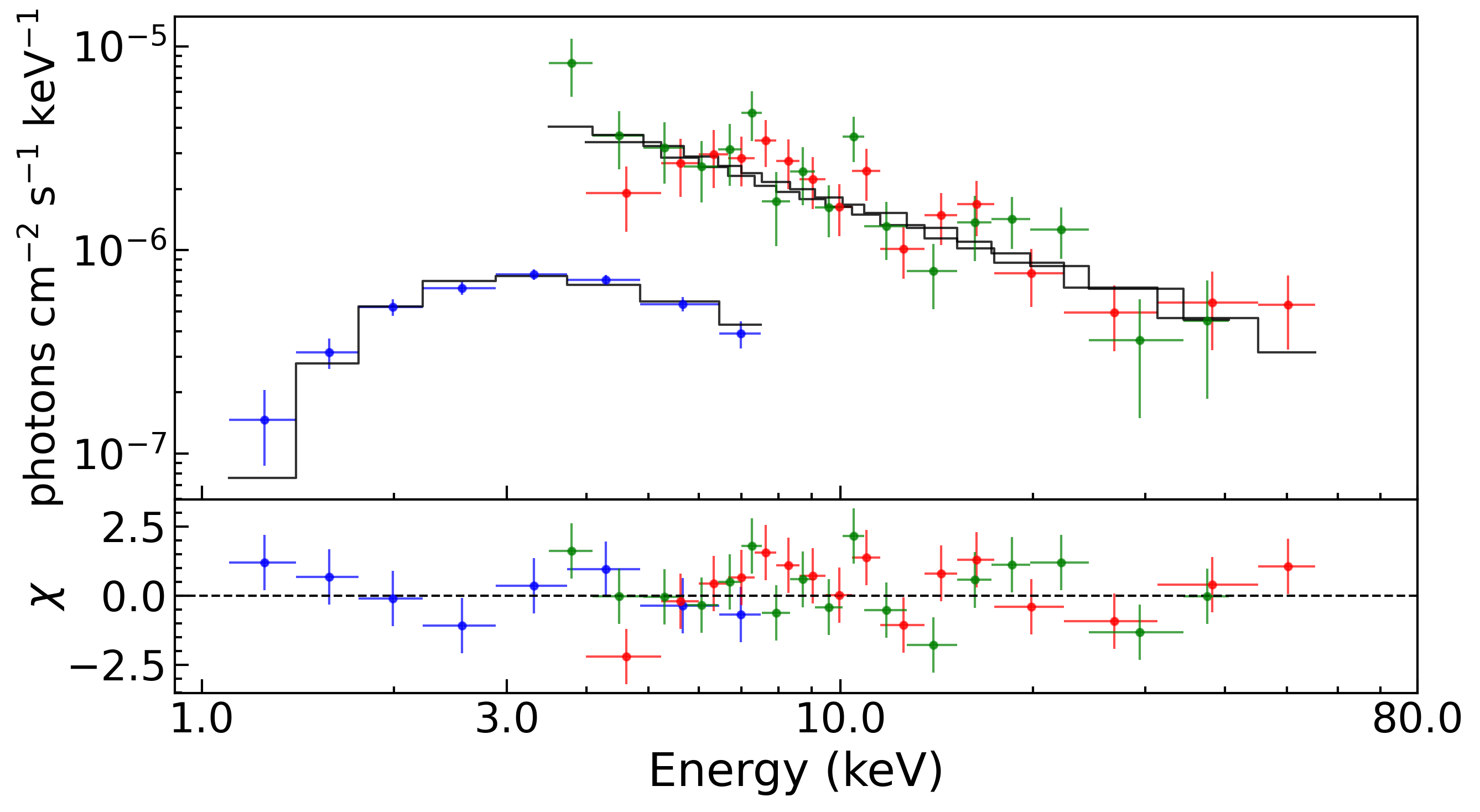}
        \caption{Inter-pulse spectra observed by NuSTAR and NICER. }
        \label{fig:spectra2_b}
    \end{subfigure}
    % 主标题
    \caption{The main pulse and inter-pulse best-fit spectra. The top panel shows the joint spectral fit of NuSTAR and NICER data with $\mathrm{N_H}$ fixed, while the bottom panel presents the residuals. The black solid lines represent the best-fit model curves for each instrument. \textbf{(a)} The blue crosses represent the NICER spectrum in the 1.2--8~keV range, the red and green crosses correspond to the NuSTAR FPMA and FPMB spectra in the 3--79~keV range respectively. \textbf{(b)} The blue crosses represent the NICER spectrum in the 1--10~keV range, the red and green crosses correspond to the NuSTAR FPMA and FPMB spectra in the 3--70~keV range respectively. All energy bands are selected to ensure that the source count rate is higher than the background.}
    \label{fig:spectra2}
\end{figure}

All spectra are well described by an absorbed power-law model (tbabs*power-law).
The best-fit results for the phase-integrated spectra are shown in Figure~\ref{fig:spectra1} and 
 Table~\ref{tab:spectra1}.
The parameters differ slightly from those obtained in \citealt{2014ApJ...790..103A}.
When fitting the main and inter-pulse spectra, we fixed the $\mathrm{N_H}$ to the best-fit value of the phase-integrated spectrum. The best-fit results for these two spectra are 
shown separately in Figure~\ref{fig:spectra2_a} and Figure~\ref{fig:spectra2_b} with the parameters listed in Table~\ref{tab:spectra1}.
Finally, we calculated both absorbed and unabsorbed flux from the best-fit model in the energy bands of 0.5--10~keV and 10--100~keV. The flux results are shown in Table~\ref{tab:spectra2}.

For the NICER data, we fitted the pulse profiles with Gaussian functions to obtain the flux ratios of the main pulse to the inter‑pulse in each energy band. The resulting main pulse to inter‑pulse flux ratios from profile fitting are as follows: the flux ratio in the 0.5--2~keV band is $1.74 \pm 0.61$, the flux ratio in the 2--4~keV band is $1.95 \pm 0.25$, the flux ratio in the 4--6~keV band is $1.79 \pm 0.28$ and the flux ratio in the 6--8~keV band is $1.70 \pm 0.48$. Simultaneously, we calculated the flux ratios using the best‑fit spectral parameters. The main pulse to inter‑pulse flux ratios are as follows: the flux ratio in the 0.5--2~keV band is $2.08^{+0.55}_{-0.40}$, the flux ratio in the 2--4~keV band is $1.99^{+0.42}_{-0.35}$, the flux ratio in the 4--6~keV band is $1.87^{+0.33}_{-0.28}$ and the flux ratio in the 6--8~keV band is $1.79^{+0.28}_{-0.24}$.
Within the measurement uncertainties, the flux ratios derived from the pulse profiles are consistent with those obtained from spectral fitting. When accounting for the measurement errors, the spectral fitting results reveal a gradual decrease in the main pulse to inter-pulse flux ratio from 0.5 keV to 8 keV. Although the profile fitting results exhibit a slight increase in the 2–4 keV band, this deviation is within the statistical uncertainties and does not contradict the overall decreasing trend indicated by the spectral analysis.

This phenomenon derives from the smaller photon index of the inter-pulse spectrum compared to the main pulse, indicating a harder inter-pulse spectrum and a larger relative contribution at higher energies compared with the main pulse.
This characteristic, observed in the Crab pulsar, can be interpreted within two physical frameworks: in the outer gap model, the harder emission of the secondary peak originates from regions closer to the neutron star surface; alternatively, particle-in-cell simulations suggest it arises from regions with a stronger accelerating electric field corresponding to the secondary peak (\citealt{2022ApJ...928..183Y}). In addition, we note that the photon indices presented in Table~\ref{tab:spectra1} are consistently higher than the values of \citealt{2014ApJ...790..103A}. Potential origins for this offset include intrinsic pulsar evolution or systematic differences in generating background spectra.

%%%%%%%%%%%%%%%%%%%%%%%%%%%%%%%%%%%%%%%%%%%%%%%%%%%%%%%%%%%%
\begin{table}
\bc
\begin{minipage}[]{1.0\textwidth}
\caption[]{The best-fit parameters of phase-integrated, main pulse and inter-pulse spectra of PSR J2022+3842\label{tab:spectra1}}\end{minipage}
\setlength{\tabcolsep}{5pt}
\small
 \begin{tabular}{llllll}
  \hline%\noalign{\smallskip}
 & Instrument& constant& $N_{H}$& $\Gamma$& $\chi^2/d.o.f$\\
  \hline\noalign{\smallskip}
 & NICER & $0.37^{+0.10}_{-0.09}$  & $2.2^{+1.0}_{-0.8}$ & $1.22(7)$ & \\
Integrated pulse & NuSTAR FPMA & $1$  & $2.2^{+1.0}_{-0.8}$ & $1.22(7)$ &  $37.71/28$\\
  & NuSTAR FPMB & $1.20^{+0.08}_{-0.07}$  & $2.2^{+1.0}_{-0.8}$ & $1.22(7)$ & \\
    \hline\noalign{\smallskip}
  %$\chi^2/d.o.f$ &  & $37.71/28$  & \\
  & NICER & $0.31(2)$  & $2.223$ & $1.17(4)$ & \\
 Main pulse & NuSTAR FPMA & $1$  & $2.223$ & $1.17(4)$  & $48.50/61$\\
 & NuSTAR FPMB & $1.06^{+0.10}_{-0.09}$  & $2.223$ & $1.17(4)$ & \\
\hline\noalign{\smallskip}
    %$\chi^2/d.o.f$ &  & $37.71/28$  & \\
  & NICER & $0.17(2)$  & $2.223$ & $1.03^{+0.07}_{-0.06}$ & \\
 Inter-pulse & NuSTAR FPMA & $1$  & $2.223$ & $1.03^{+0.07}_{-0.06}$ &  $41.86/36$\\
 & NuSTAR FPMB & $1.08^{+0.13}_{-0.12}$  & $2.223$ & $1.03^{+0.07}_{-0.06}$ & \\
 \hline%\noalign{\smallskip}
\end{tabular}
\ec
%% place \tablecomments and \tablerefs below \end{center| and \end{center}:
%% you may leave the table-width parameter to editors or set to your actual size
\tablecomments{0.86\textwidth}{Table 3 shows the best-fit values of the phase-integrated, main pulse and inter-pulse spectra. The unit of $\mathrm{N_H}$ is $\mathrm{10^{22}~\text{cm}^{-2}}$.}
\end{table}
%%%%%%%%%%%%%%%%%%%%%%%%%%%%%%%%%%%%%%%%%%%%%%%%%%%%%%%%%%
\begin{table}
\bc
\begin{minipage}[]{1.0\textwidth}
\caption[]{Flux derived from the best-fit parameters of phase-integrated, main pulse and inter-pulse spectra of PSR J2022+3842\label{tab:spectra2}}\end{minipage}
\setlength{\tabcolsep}{5pt}
\small
 \begin{tabular}{lllll}
  \hline%\noalign{\smallskip}
 & $F^{abs}_{0.5-10keV}$& $F^{abs}_{10-100keV}$& $F^{unabs}_{0.5-10keV}$& $F^{unabs}_{10-100keV}$\\
  \hline\noalign{\smallskip}
Integrated pulse & $7.4(5)$ & $58^{+8}_{-7}$  & $8.9(6)$ & $59^{+8}_{-7}$\\
    \hline\noalign{\smallskip}
  %$\chi^2/d.o.f$ &  & $37.71/28$  & \\
Main pulse & $27(2)$ & $245^{+22}_{-21}$  & $33.2(2)$ & $245^{+22}_{-21}$\\
\hline\noalign{\smallskip}
    %$\chi^2/d.o.f$ &  & $37.71/28$  & \\
Inter-pulse & $25(5)$ & $292^{+40}_{-36}$  & $29(3)$ & $292^{+18}_{-36}$\\
  \hline%\noalign{\smallskip}
\end{tabular}
\ec
%% place \tablecomments and \tablerefs below \end{center| and \end{center}:
%% you may leave the table-width parameter to editors or set to your actual size
\tablecomments{0.86\textwidth}{Table 4 shows the flux calculated using the best-fit values of the phase-integrated, main pulse and inter-pulse spectra. The unit of all flux is $\mathrm{10^{-13}~\text{erg}~\text{cm}^{-2}~\text{s}^{-1}}$.}
\end{table}

\section{Discussion and Conclusions} \label{sec:discuss}

This pulsar exhibits X-ray radiation characteristics consistent with the Crab Pulsar -- the X-ray spectra can be well fitted by a single power-law model and the pulse profile is stable after large glitches. However, its glitch activities resemble more closely with the Vela Pulsar -- the activity parameter is larger than that of the Crab Pulsar and no small glitches have been detected. Given that the pulsar's characteristic age is 8.9~kyr, which places it at the boundary between Crab-like and Vela-like pulsars, long-term continuous timing and spectral observations may reveal how its radiation, internal and external structure, and coupling mechanisms evolve as it slows down and cools (\citealt{2025SCPMA..6819505G, 2025SCPMA..6819503L, 2025SCPMA..6819502Z}).

\begin{figure}[htbp]
    \centering
    \includegraphics[width=20cm, height=10cm, keepaspectratio]{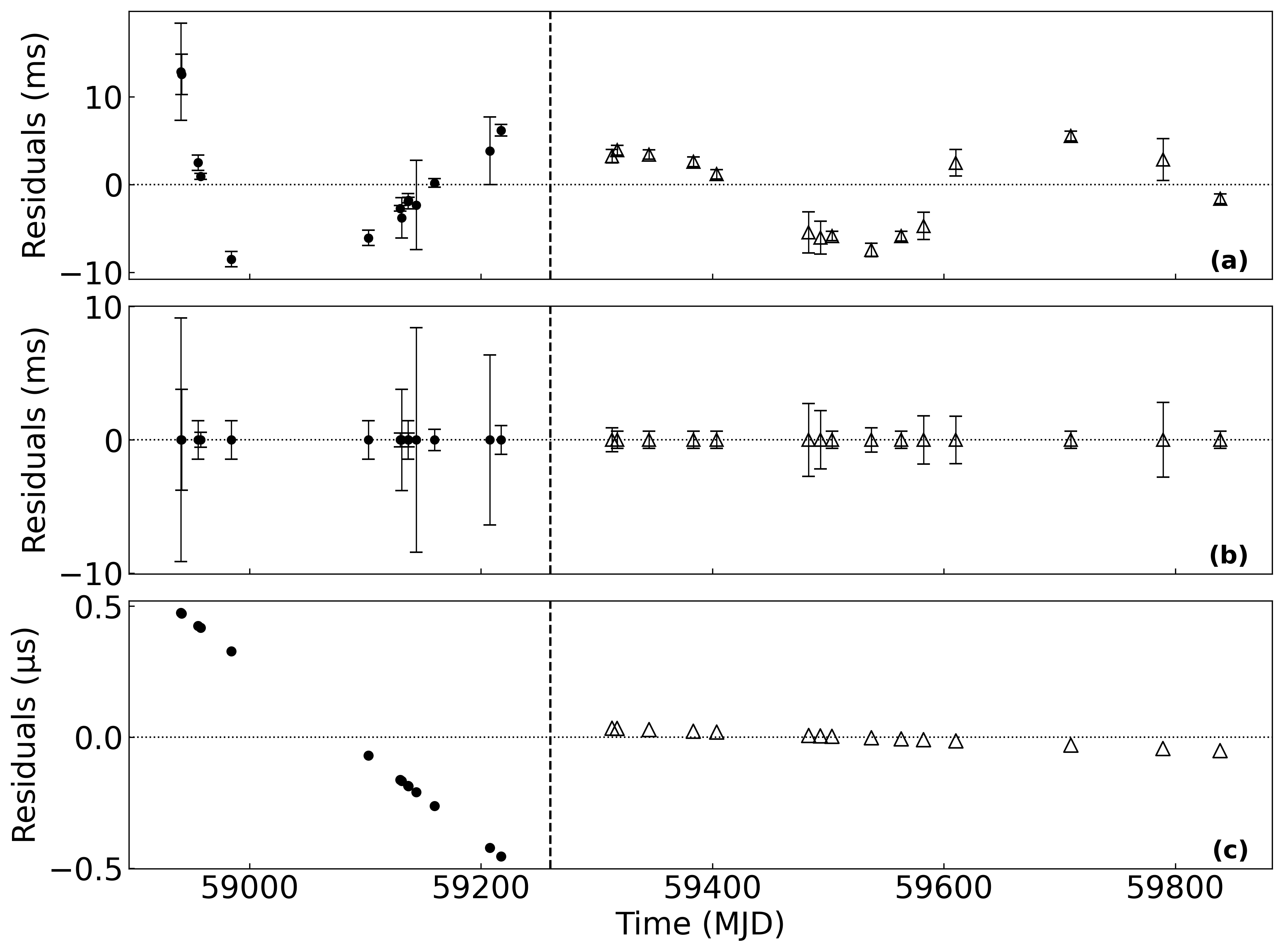}
    \caption{\textbf{(a)} Post-fit timing residuals using the ephemeris of segment 3 and ToAs from segment 3 (filled black circle) and segment 4 (open black triangle). \textbf{(b)} Timing residuals with $\mathrm{1\sigma}$ error bars after subtracting the impact of timing noise calculated by ENTERPRISE. The filled black circles present data before MJD 59260 and the open black triangles present data after MJD 59260. \textbf{(c)} Timing residuals without error bars after subtracting the impact of timing noise calculated by ENTERPRISE. The filled black circles present data before MJD 59260 and the open black triangles present data after MJD 59260. The verticle black dashed lines in MJD 59260 indicate the epoch when the ephemeris changes. No significant evidence for a small glitch event was found within the $\mathrm{1\sigma}$ confidence interval.}
    \label{fig:residual_all}
\end{figure}

Following the method described in~\ref{subsec:timing}, we calculated the BF factor for both segments and used a white noise model to describe the timing noise during MJD 58875--59260 and the red noise plus white noise model with 12 modes for MJD 59260--59900. The timing residuals obtained using the ephemeris of segment 3 and the ToAs of segments 3 and 4 in tempo2, as well as the residuals after subtracting the timing noise calculated by ENTERPRISE are shown in Figure~\ref{fig:residual_all}.
The results of PTMCMC are shown in Figure~\ref{fig:noise1_combined}.
The residuals derived from tempo2 and ENTERPRISE noise models are shown in Figure~\ref{fig:residual_combined}. Whether through visual inspection or subtracting the impact of timing noise, we can conclude that no glitch event occurred around MJD 59260 within the $\mathrm{1\sigma}$ error range. The strong timing noise is likely the result of the recovery process from the large glitch around MJD 58875.

The red noise affects the value of the second frequency derivative, leading to an inevitable overestimation of it during the pulsar timing analysis in tempo2. When using ENTERPRISE to analyze the tempo2 output, the portion of noise "absorbed" by the second frequency derivative is no longer modeled, and its effect is not subtracted from the timing residuals. Consequently, after removing other known contributions of timing noise calculated in ENTERPRISE, the remaining residuals exhibit a systematic downward trend. We suggest that this trend is caused by the second frequency derivative, which is contaminated by the underlying red noise. Mathematically, this trend can be effectively removed using a cubic function of time. Within the time span of our study, a quadratic function also provides a satisfactory fit.
\FloatBarrier

\normalem
\begin{acknowledgements}
We gratefully acknowledge the use of public data from the NICER, NuSTAR archives and the codes from ENTERPRISE, DYNESTY and PTMCMCSampler. This work is also supported by the Natural Science Foundation of China (12233006, 12373046), 
the Foundations from Yunnan Province (202301AS070073) and the China's Space Origins Exploration Program and the National Natural Science Foundation of China (Grant Nos. 12373051).
\end{acknowledgements}

%%%%%%%%%%%%%%%%%%%%%%%%%%%%%%%%%%%%%%%%%%%%%%%%%%%%%%%%%%%%%%%%%%%%%%%%%%

\bibliographystyle{raa}
\bibliography{bibtex}

\appendix
\section{PTMCMC Result}

\begin{figure}[htbp]
    \centering
    % 子图 (a)
    \begin{subfigure}[t]{0.40\textwidth}
        \centering
        \includegraphics[width=8cm, height=5cm, keepaspectratio]{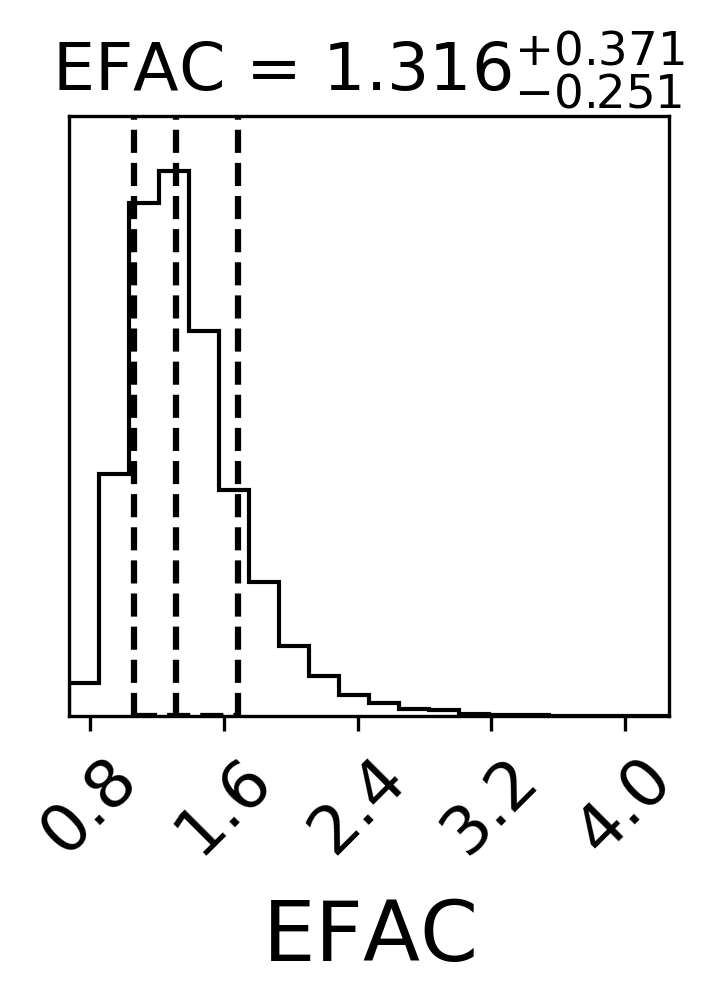}
        \caption{Best-fit white noise model parameter for MJD 58875--59260} % 子图标题
        \label{fig:noise1_a}
    \end{subfigure}
    \hfill
    % 子图 (b)
    \begin{subfigure}[t]{0.58\textwidth}
        \centering
        \includegraphics[width=8cm, height=9cm, keepaspectratio]{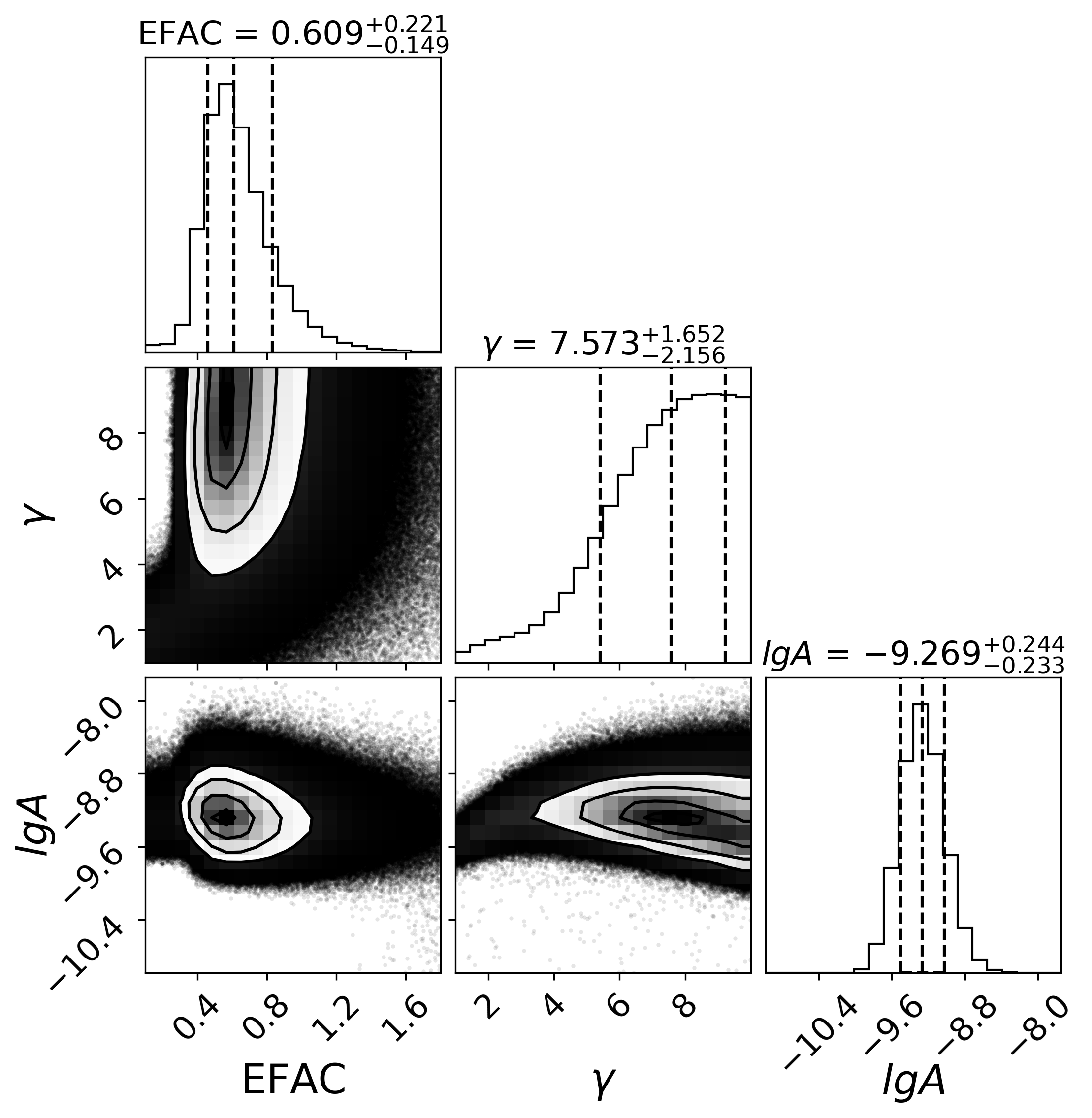}
        \caption{Best-fit white and red noise model parameters for MJD 59260--59900} % 子图标题
        \label{fig:noise1_b}
    \end{subfigure}
    % 主标题
    \caption{The results of PTMCMC for both segments. Due to the lack of data, the $\mathrm{\gamma}$ result exhibits a long tail extending beyond $\mathrm{\gamma=10}$ between $\mathrm{1\sigma}$ and $\mathrm{2\sigma}$. Therefore we only plotted the portion where $\mathrm{\gamma}$ is less than 10.}
    \label{fig:noise1_combined}
\end{figure}

\begin{figure}[htbp]
    \centering
    % 子图 (a)
    \begin{subfigure}[c]{1.0\textwidth}
        \centering
        \includegraphics[width=14cm, height=8cm, keepaspectratio]{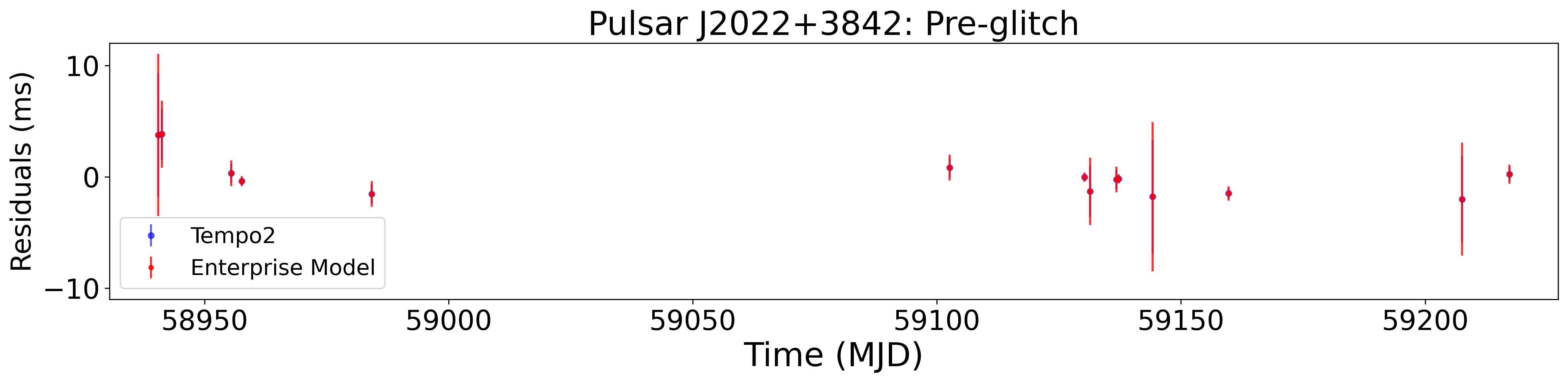}
        \vspace{-7pt}  % 图像与 caption 距离调近
        \caption{Original and calculated timing residuals during MJD 58875--59260.} % 子图标题
        \label{fig:residual1_a}
    \end{subfigure}
    \\[8pt]
    %\hfill
    % 子图 (b)
    \begin{subfigure}[c]{1.0\textwidth}
        \centering
        \includegraphics[width=14cm, height=8cm, keepaspectratio]{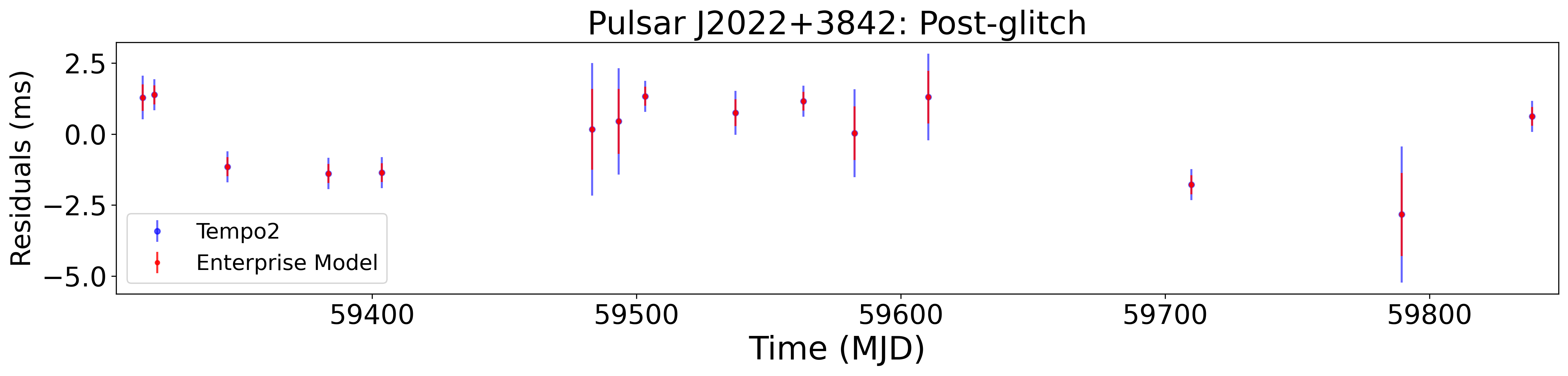}
        \vspace{-7pt}  % 图像与 caption 距离调近
        \caption{Original and calculated timing residuals during MJD 59260--59900.} % 子图标题
        \label{fig:residual1_b}
    \end{subfigure}
    % 主标题
    \caption{The residuals from tempo2 and noise modeling.  The blue dots with error bars are residuals derived from tempo2, the red dots with error bars correspond to the residuals calculated using the ephemeris timing model plus the white noise model, with parameter value from Figure~\ref{fig:noise1_a} and Figure~\ref{fig:noise1_b}.}
    \label{fig:residual_combined}
\end{figure}

The figures of the result of PTMCMC and the origional timing residuals with the residuals calculated by each noise model are shown here in Figure~\ref{fig:noise1_combined} and Figure~\ref{fig:residual_combined}.

%%%%%%%%%%%%%%%%%%%%%%%%%%%%%%%%%%%%%%%%%%%%%%%%%%%%%%%%%%%%%%%%%%%%%%%%%%
Facilities we worked on: NICER (XTI), NuSTAR (FPMA and FPMB).
Software used in our work: HEASoft (v6.34) (\citealt{2014ascl.soft08004N}), tempo2 (\citealt{2006MNRAS.369..655H}), ENTERPRISE (\citealt{2020zndo...4059815E})
DYNESTY (A Dynamic Nested Sampling Package for Estimating Bayesian Posteriors and Evidence) (\citealt{2020MNRAS.493.3132S, 2004AIPC..735..395S, 10.1214/06-BA127, Higson2018, 2016S&C....26..383B, 2019PASP..131j8005B, 2023zndo...3348367K})
PTMCMC, data sampler and parameter estimater (\citealt{2017zndo...1037579E})

\end{document}